\documentclass[pre,twocolumn,aps,showpacs]{revtex4}

\usepackage{graphicx}% Include figure files
\usepackage{dcolumn}% Align table columns on decimal point
\usepackage{bm}% bold math

\begin{document}
\title{

Extreme events in discrete nonlinear lattices

}
\author{
A. Maluckov$\ ^{1}$, Lj. Had{\v z}ievski$\ ^{2}$,
N. Lazarides$\ ^{3,4}$, and G. P. Tsironis$\ ^{3}$
}
\affiliation{
$\ ^{1}$Faculty of Sciences and Mathematics, Department of Physics,
P. O. Box 224, 18001 Ni{\v s}, Serbia \\
$\ ^{2}$Vin{\v c}a Institute of Nuclear Sciences,
P. O. Box 522, 11001 Belgrade, Serbia \\
$\ ^{3}$Department of Physics, University of Crete,
and
Institute of Electronic Structure and Laser,
Foundation for Research and Technology -- Hellas,
P. O. Box 2208, 71003 Heraklion,  Greece \\
$\ ^{4}$Department of Electrical Engineering,
Technological Educational Institute of Crete,
P. O. Box 140, Stavromenos, 71500, Heraklion, Crete, Greece
}
\date{\today}
\begin{abstract}
We perform statistical analysis on discrete nonlinear waves
generated though modulational instability in the context of the
Salerno model that interpolates between the intergable
Ablowitz-Ladik (AL) equation and the  nonintegrable discrete
nonlinear Schr\"odinger (DNLS) equation.  We focus on extreme
events in the form of discrete rogue or freak waves that may arise
as a result of rapid coalescence of discrete breathers or other
nonlinear interaction processes.  We find  power law dependence in
the wave amplitude distribution accompanied by an enhanced
probability for freak events close to the integrable limit of the
equation. A characteristic peak in the extreme event probability
appears that is attributed to the onset of interaction of the
discrete solitons of the AL equation and the accompanied
transition from the local to the global stochasticity monitored
through the positive Lyapunov exponent of a nonlinear map.

\end{abstract}

%%\preprint{APS/123-QED}

\maketitle {\em Introduction.-} The motivation of the present work
stems from  observations  of the sudden appearance of extremely
large amplitude sea waves referred to as rogue or freak waves
\cite{Kharif}.  These waves appear very suddenly in relatively
calm seas, reach amplitudes of over $20m$ and may destroy or sink
small as well as large vessels \cite{Muller}.  Theoretical
analysis of ocean freak waves has been linked to nonlinearities in
the waver wave equations, studied though the nonlinear
Schr\"odinger (NLS) equation and shown that the probability of
their appearance is not insignificant \cite{Onorato}.  A scenario
for freak wave generation in NLS is through a Benjamin-Feir
(modulational) instability, resulting in self-focusing effects and
subsequent formation of freak waves \cite{Zakharov}. Modulational
instability (MI) induces local exponential growth in the wave
train  amplitude \cite{Onorato1,Shukla}  that has been confirmed
experimentally and numerically \cite{Ruban}.

Intriguingly, there are completely different physical systems that
possess the required nonlinear characteristics which favour the
appearance of rogue waves. Recent  observation of optical rogue
waves in a microstructured optical fiber was reported \cite{Solli}
in  a regime near the threshold of soliton-fission supercontinuum
generation, i.e., in a region where MI plays a key role in the
dynamics. A generalized NLS equation was used successfully to
model the generation of optical rogue waves while, additionally,
control  and manipulation of rogue soliton formation was also
discussed \cite{Dudley}. The mechanism of the rogue waves
creation, or, more generally   of extreme events,
 has become an issue of principal interest
in various other contexts  as well, since rogue waves can signal
catastrophic phenomena such as an earthquake, a thunderstorm, or a
severe financial crisis. Knowledge of the probability of
occurrence of extreme events and the capability to predict the
time at which such an event may take place is of a great value.
Such events are usually rare, and they exhibit "extreme-value"
statistics, typically characterized by heavy-tailed probability
distributions. Experimental observation of optical rogue-wave-like
fluctuations in fiber Raman amplifiers show that the probability
distribution of their peak power follows a power law
\cite{Hammani}.

In this work we focus on the discrete counterparts of rogue waves
that may appear in nonlinear lattices as a result of discrete
soliton or breather induction and their mutual interactions.
Specifically we investigate the role of integrability  in the
formation of discrete rogue waves (DRW) and the resulting extreme
event statistics. Their appearance may affect dramatically the
physical systems.  We use the Salerno model  \cite{Salerno}  that
through a unique parameter interpolates between a fully integrable
discrete lattice, viz.  the Ablowitz-Ladik (AL) lattice
\cite{Ablowitz} , and  the nonintegrable  DNLS equation
\cite{ELS,MT}. One of the basic questions to be addressed below is
the probability of occurrence of a DRW as a function of the degree
of integrability of the lattice and thus study the role of the
latter in the production of extreme lattice events \cite{Nicolis}.

{\em The Salerno model.-}
The Salerno model (SM) is given through the following set of equations
\begin{eqnarray}
\label{eq1}
   i\frac{d\psi_n}{dt}=
   -(1+\mu |\psi_n|^2)(\psi_{n+1}+\psi_{n-1})
  -\gamma |\psi_n|^2\psi_{n}
\end{eqnarray}
where  $\mu$ and $\gamma$ are two nonlinearity parameters. When
$\mu =0$ the model becomes the DNLS equation while for $\gamma =0$
it reduces to AL.  Several properties of the model such as
integrability \cite{Rumpf} and stability of localized travelling
waves \cite{Cai,Hennig} have been analyzed. Both  the norm $N$ and
the Hamiltonian $H$ of the model are conserved quantities. They
are given by
\begin{eqnarray}
   N&=&\frac{1}{\mu}\sum_n \ln{|1+\mu|\psi_n|^2|},  \label{eq2} \\
   H&=&\sum_n \left[\frac{\gamma}{\mu^2} \ln|1+\mu|\psi_n|^2|
    -\frac{\gamma}{\mu} |\psi_n|^2 -2 Re[\psi_n\psi_{n+1}^*]  \right].
\label{eq3}
\end{eqnarray}
It is also known that Eq. (\ref{eq1}) exhibits MI, which may give
rise to stationary, spatially localized solutions in the form of
discrete breathers (DBs), i.e., periodic and spatially localized
nonlinear excitations \cite{DBs}. The MI induced DBs appear in
random  lattice locations  and may be mobile. High-amplitude DBs
tend to absorb low-amplitude ones, resulting after some time in a
small number of very high amplitude excitations, which may get
pinned at a specific lattice site due to the Peierls-Nabarro
potential barrier in nonintegrable lattices \cite{Kivshar1}. In
general, high-amplitude DBs are virtual bottlenecks which slow
down the relaxation processes in  nonlinear lattices
\cite{Tsironis,Rasm}, and it has been proposed that they may serve
as models for freak waves \cite{Dysthe}. The development of MI in
Eq. (\ref{eq1}) can be analyzed with the linear stability analysis
of its the plane wave solutions perturbed by small phase and
amplitude perturbations \cite{Maluckov}. The interplay between the
on-site and intersite nonlinear terms (i.e., according to the
variation of their relative strength through $\mu$ and $\gamma$),
may change MI properties and, consequently, the conditions for the
DBs to exist in the lattice \cite{Kivshar}. The SM has recently
found applications in modelling Bose-Einstein condensates of
dipolar atoms in a strong periodic potential \cite{Gomez}, dilute
Bose-Einstein condensates trapped in a periodic potential
\cite{Trombettoni}, and even biological systems \cite{Salerno1}.

For later convenience in the numerical simulation, the variables
$\psi_n$ in Eq. (\ref{eq1}) are rescaled as
$\psi_n=\xi_n/\sqrt{\mu}$, so that in terms of $\xi_n$ the dynamic
equations read
\begin{eqnarray}
\label{eq4}
  i\frac{d\xi_n(t)}{dt}=
    -(1 +|\xi_n(t)|^2)(\xi_{n+1}+\xi_{n-1})
    -\Gamma |\xi_n(t)|^2\xi_{n} ,
\end{eqnarray}
where $\Gamma=\gamma/\mu$. Therefore, the whole two-dimensional
parameter space $(\gamma,\mu)$ can be scaled by $\mu=1$,
leaving  $\gamma$ as a free parameter. With that scaling
we may go as close to the DNLS limit as we want to,
by simply let $\Gamma$ to attain very large values.
However, the exact DNLS limit $\mu=0$ has to be calculated
separately.
%%******************figure1********************************************
\begin{figure}[h]
\includegraphics[angle=0, width=0.8\linewidth]{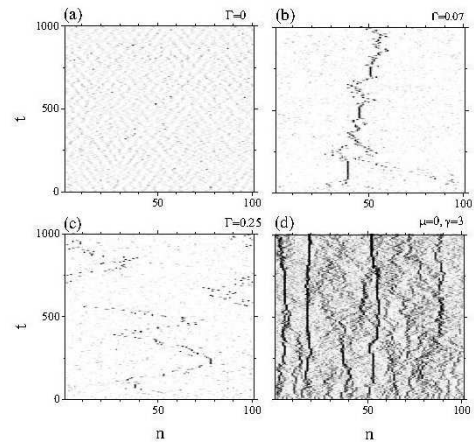}
\caption{\label{Fig. 1} Evolution of the scaled amplitudes
$|\xi_n|$ for a lattice of size $N=101$, with $\Gamma$ ($\mu$ and
$\gamma$ in the DNLS case), is shown on the figure. The initial
conditions for all cases are $\xi_n=1$ for any $n$ (uniform) plus
a small amount of white noise. }
\end{figure}
%%**********************************************************************

{\em Statistics of extreme events.-} We integrate numerically the
system of Eqs. (\ref{eq4}) with periodic boundary conditions using
a sixth order Runge-Kutta algorithm with fixed time-stepping
$\Delta t=10^{-4}$. We started simulations with different initial
conditions (the plane wave, uniform background with white noise
and Gaussian noise) which gave similar results. Here we present
calculations in which the initial condition is uniform, $\xi_n =1$
for any $n$, with the addition of a small amount of white noise to
accelerate the development of the MI. The uniform solution is
chosen in the interval where it is known from linear stability
analysis that it is unstable. By varying the nonlinearity
parameters we identify broadly three regimes of DRWs that are
shown as spatiotemporal evolutions in Fig. \ref{Fig. 1}. For the
purely integrable AL lattice ($\Gamma=0$) DBs are mobile and
essentially noninteracting; as a result we do not observe
significant formation of high DRWs (Fig. \ref{Fig. 1}a). In the
vicinity of the AL limit, i.e. for  small $\Gamma$ ($\Gamma \sim
0.1$),  there is an onset of weak interaction of the localized
modes of the AL lattice leading to a significant increase in DRW
formation that are mobile (Fig. \ref{Fig. 1}b,c). In this regime
the DBs are highly mobile indicating that DB merging could be
responsible for creation of high-amplitude localized waves. For
$\Gamma >> 0.1$ on the other hand, DNLS-type behavior dominates
the SM and localized structures that are initially created through
the MI become easily trapped in the lattice (Fig. \ref{Fig. 1}d).

\begin{figure}[!t]
\includegraphics[angle=0, width=0.8\linewidth]{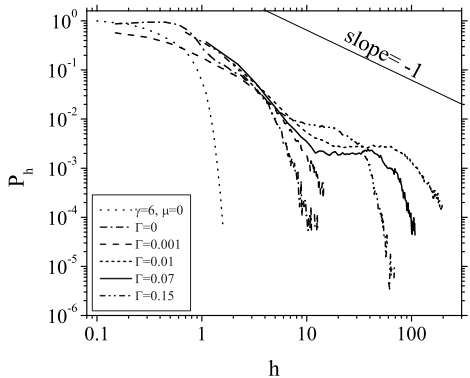}
\caption{\label{Fig. 2} The normalized height probability density
$P_h (h)$ for several values of $\Gamma$ and for the DNLS limit
(with $\gamma=6$). The line with slope $-1$ is added to assist
comparisons and corresponds to $P_h \sim 1/h$. Approximately
vertical drop corresponds to the DNLS limit with an exponential
tail. The increase of $\Gamma (\mu=1)$ leads to the decrease of
the slope and appearance of plateau on the $P_h$ curve; the latter
increases the extreme event probability leading a maximum at
 $\Gamma=0.07$ (Fig. 3.) }
\end{figure}

The three regimes mentioned previously are probed by calculating
the time-averaged height distributions $P_h$. We first define the
forward (backward) height at the $n-$th site as  the difference
between two successive minimum (maximum) and maximum (minimum)
values of $|\xi_n (t) |$. We use then both the forward and the
backward heights for the calculation of the local height
distribution; after spatial averaging the latter results in the
height probability densities (HPDs) shown in Fig. 2.  We note that
the tails of the HPDs are related to extreme events and the
appearance of DRWs. For $\Gamma$ finite the HPDs are sharply
peaked but have extended tails indicating that extreme events are
more than several times as large as the  mean distribution height.
In the DNLS limit ($\Gamma \gg 1$) the obtained HPD is very close
to the Rayleigh distribution whose tails decay very fast
\cite{VanKampen}, indicating negligible probability for the
occurrence of extreme events (dotted curve in Fig. \ref{Fig. 2}).
In all the other cases the decay of the tails of the HPDs is much
slower.

\begin{figure}[t]
\includegraphics[angle=0, width=0.8\linewidth]{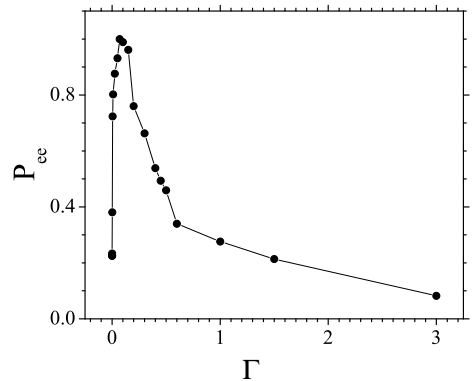}
\caption{\label{Fig. 3} The normalized probability $P_{ee}=P_h
(h\ge h_{th})$ for the occurrence of extreme events as a function
of the integrability parameter $\Gamma$. All data present averaged
results of five numerical measurements differing in the initial
conditions. }
\end{figure}

In order to probe further the onset of extreme discrete events we employ the practice used in water waves and define
a DRW as one that has a height greater than
 $h_{th}=2.2 h_s$,  with $h_s$ being the significant  wave height.
The latter is defined as the average height of the one-third
higher waves in the height distribution. As a result, the
probability of occurrence of extreme DRW events $P_{ee} = P_h
(h>h_{th})$ is obtained by integration of the (normalized) HPD
from $h=h_{th}$ up to infinity.    By evaluating several HPDs as
those in Fig. \ref{Fig. 2} we may estimate the probability of
occurrence of DRWs $P_{ee}$ as a function of the parameter
$\Gamma$ (the results are shown in Fig. \ref{Fig. 3}).  We note
that the probability for the occurrence of a DRW  has a certain
value in the AL case, subsequently peaks for small values of
$\Gamma$ and decays precipitously when  $\Gamma >>1$.  This
behavior of the probability $P_{ee}$ is compatible with the DB
picture outlined earlier, viz. in the very weakly nonintegrable
regime the AL modes may interact leading to DB fusion and DRW
generation.  On the other hand,  as nonintegrability becomes
stronger,  the scattering of the AL modes is  more chaotic leading
to a suppression of DRW formation.

%%******************figure2********************************************
%%*********************************************************************

%%******************figure3********************************************
%%*********************************************************************

{\em Map approach.-} In order to probe deeper on the formation of
DRWs we substitute $\psi_n=\phi_n \exp(-i\omega t)$ into Eq.
(\ref{eq1}), with $\phi_n$ a real-valued function of the lattice
site $n$, and obtain the stationary equation
\begin{eqnarray}
  \label{eq5}
   \omega\phi_n+(1+\mu |\phi_n|^2)(\phi_{n+1}+\phi_{n-1})
     +\gamma |\phi_n|^2\phi_{n}=0 ,
\end{eqnarray}
which can be transformed in the two-dimensional map
\begin{eqnarray}
  \label{eq6}
     x_{n+1}=-\frac{\omega+\gamma x_n^2}{1+\mu x_n^2}x_n-y_n ,
    \qquad  y_{n+1}=x_n ,
\end{eqnarray}
where we have defined $x_n=\phi_n$ and $y_n=\phi_{n-1}$.
Eqs. (\ref{eq6}) represent a real analytic area-preserving map
\cite{Hennig,Hennig1} with the lattice index $n$ playing the role of
discrete 'time'.

\begin{figure}[!t]
\includegraphics[angle=0, width=0.7\linewidth]{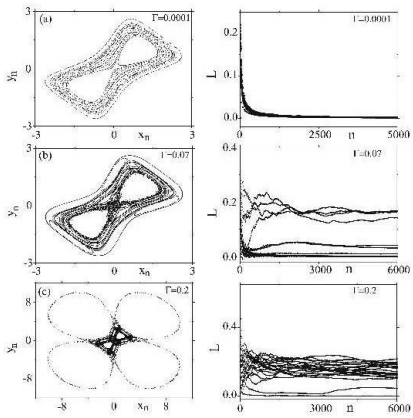}
\caption{\label{Fig. 4} Orbits started at different initial
positions in the neighborhood of map origin and corresponding the
one-dimensional Liapunov exponents. }
\end{figure}

The phase portraits of the map  Eq. (\ref{eq6}) for several
$\Gamma$-values are shown in Fig. \ref{Fig. 4}.  In the AL limit,
the phase space consists of perfectly disconnected separatrices
while for non-zero $\Gamma$, the stable and unstable manifolds
intersect transversely, resulting in the generation of a
homoclinic tangle. With increasing $\Gamma$ the motion near
separatrices becomes exceedingly complicated and  the trajectories
wander irregularly before approaching an attracting set (Figs. 4b
and 4c). Moreover, for any $\Gamma \ne 0$, the position of
separatrices in phase space changes in time, resulting in
overlapping of neighboring separatrices and diffusion in those
regions which have been traversed by a separatrix. The sharp peak
of the probability of occurrence of extreme events $P_{ee}
(h>h_{th})$ in the SM (Fig. \ref{Fig. 3}) can be associated  with
the opening of a stochasticity web, when orbits fast explore all
extended narrow stochasticity regions leading to an anomalous
relaxation phase \cite{Rumpf,VanKampen}. This event signs the
transition from the local to global stochasticity
\cite{Lichtenberg} in SM. On the other hand, the decrease of
$P_{ee} (h>h_{th})$ for larger $\Gamma$'s is related to the
increasingly longer trapping time in more developed stochasticity
region.

The Melnikov analysis in the SM \cite{Hennig} shows that the
magnitude of the separatrix splitting and the consequent
development of stochasticity depends on the $\Gamma/|\omega|$
ratio. The conjecture that $P_{ee}$ is associated with the
complexity of the phase portraits of the corresponding maps
implies that $P_{ee}$ should also depend on the $\Gamma/|\omega|$
ratio. In our case $|\omega|$ is related to the modulation
frequency of the initially uniform solution $U$ with the relation
$|\omega|=(\gamma+2\mu)U^2 +2$, which, through the MI process it
transformed into a train of localized DB-like configurations. We
have checked numerically that for fixed ratio $\Gamma/|\omega|$
and different values of $U$ and $\Gamma$ we obtain the same HPD.
As a consequence, the probability of extreme events $P_{ee}$ as a
function of the $\Gamma/|\omega|$ is qualitatively the same with
that of $P_{ee}$ as a function of $\Gamma$ shown in Fig. \ref{Fig.
3}.

The degree of nonintegrability in the SM model can be quantified
by calculating  the Lyapunov exponents of the corresponding maps
\cite{Maluckov1}. We have thus calculated the maximum Lyapunov
exponent $L$ \cite{Lichtenberg} for the map Eq. (\ref{eq6}), for
the parameters used in the calculation of the phase portrait shown
in the left panels of Fig. \ref{Fig. 4}. It is observed that
homoclinic orbits which correspond to perfect separatrices are
characterized by vanishing Lyapunov exponent (Fig. \ref{Fig. 4}a).
With increasing stochasticity, $L$ tends to a finite positive
value which generally depends on the values of the parameters and
the initial conditions (Figs. \ref{Fig. 4}a and \ref{Fig. 4}b).

%%******************figure4********************************************
%%**********************************************************************

{\em Conclusions.-} The probability of occurrence of extreme
events $P_{ee}$ in the SM results from the competition between the
self-focusing and the energy transport mechanisms which are
implicitly correlated with the degree of integrability of the
model \cite{Rumpf}. Through modulational instability and starting
from a slightly perturbed uniform background we can generate
high-amplitude localized moving structures of the DB type that
lead to the formation of extreme events of DRW type. Depending on
their number, amplitude and life-time, they may prevent of
facilitate the energy flow in the lattice, affecting thus the
probability of extreme event formation $P_{ee}$. We find that the
latter probability depends strongly on  $\Gamma$ that affects the
degree of integrability of the lattice: DRW are much more probable
very close to the integrable SM limit rather than in the
nonintegrable one.  We find a resonance-like maximum in $P_{ee}
(\Gamma )$ that, through a nonlinear map approach, is linked to
separatrix breaking and the onset of global stochasticity. This
regime corresponds physically to weak interaction between the
quasi-integrable modes of the system.

A. M. and Lj.H. acknowledge support from the Ministry of Science
of Serbia (Project 141034). One of us (GPT) acknowledges discussions with Oriol Bohigas.

\end{document}